%% file: main.tex
\documentclass{article}
\usepackage{spconf}
\usepackage{subcaption}
\usepackage{amsmath,amssymb}
\usepackage{hyperref}
\usepackage{url}
\usepackage{booktabs}       
\usepackage{amsfonts}       
\usepackage{nicefrac}       
\usepackage{microtype}      
\usepackage{xcolor}         
\usepackage{multicol}
\usepackage{lipsum}
\usepackage[pdftex]{graphicx}
\usepackage[numbers, sort]{natbib}
\usepackage{tikz}
\usepackage{wrapfig}
\usepackage{pgfplots, pgfplotstable}
\pgfplotsset{compat=1.18}
\usetikzlibrary{calc, shapes}

\makeatletter
\DeclareRobustCommand\onedot{\futurelet\@let@token\@onedot}
\def\@onedot{\ifx\@let@token.\else.\null\fi\xspace}

\makeatother

\title{Removing Speaker Information from Speech Representation using Variable-Length Soft Pooling}
\name{Injune Hwang$^{\star}$, Kyogu Lee$^{\star}$}
\address{
  $^{\star}$Music \& Audio Research Group (MARG), Seoul National University, Republic of Korea
}

\begin{document}

\maketitle
\begin{abstract}
Recently, there have been efforts to encode the linguistic information of speech using a self-supervised framework for speech synthesis. However, predicting representations from surrounding representations can inadvertently entangle speaker information in the speech representation. This paper aims to remove speaker information by exploiting the structured nature of speech, composed of discrete units like phonemes with clear boundaries. A neural network predicts these boundaries, enabling variable-length pooling for event-based representation extraction instead of fixed-rate methods. The boundary predictor outputs a probability for the boundary between 0 and 1, making pooling soft. The model is trained to minimize the difference with the pooled representation of the data augmented by time-stretch and pitch-shift. To confirm that the learned representation includes contents information but is independent of speaker information, the model was evaluated with libri-light's phonetic ABX task and SUPERB's speaker identification task.
\end{abstract}


\input{01_introduction}

\input{02_proposed_framework}

\input{03_experiment}

\section{Conclusion}
Using the proposed soft pooling module has the effect of increasing phonetic information while concurrently diminishing speaker information within the speech representation for limited resources. 
In the context of self-supervised speech representation learning, the boundary prediction has proven advantageous in acquiring disentangled content-related information.
An intriguing observation arises from the fact that the boundaries identified as optimal for the objective function closely align with the phoneme boundaries.
Our strategy entails enhancing the model's capacity and implementing both quantization techniques and a unit-language model.

\vfill\pagebreak

\bibliographystyle{IEEEbib}
\bibliography{main}

\end{document}

%% file: 01_introduction.tex
\section{Introduction}

Numerous self-supervised speech representation learning methods, including Contrastive Predictive Coding(CPC) \cite{oord2018representation}, HuBERT \cite{hubert}, and wav2vec2.0 \cite{wav2vec2} aim to encompass general information in speech representations, such as voice characteristics, semantic content, and emotional nuiances.
While these representations offer distinct advantages in enabling diverse downstream speech-related applications, it remains crucial to extract the specific content information within them. This need is particularly prominent when sharing the embedding space with other modals that have unrelated voice characteristics. Examples of this include matching speech representations from text for text-to-speech tasks or reconstucting speech from signals from EMG sensors attached to the face in the field of silent speech \cite{scheck2023multi}. Moreover, the disentanglement of speech representations can alleviate the load associated with spoken language modeling \cite{lakhotia-etal-2021-generative} and speech synthesis \cite{choi2021neural} by reducing the non-linguistic information that is noisy to other modules.

To disentangle non-linguistic information, we will employ a prior assumption that speech is constructed from discrete linguistic units, which are discernible through distinct boundaries.
\cite{cuervo2022variable} and \cite{dieleman2021variable} represent instances of a variable-rate downsampling technique, which enables the model to leverage the observation that the informational content of speech is distributed non-uniformly.
To employ this prior more clearly, there are also methods that quantize the representation by k-means clustering and merge identical codes.(\cite{nguyen2020zero, lakhotia-etal-2021-generative}) 
While these methods are capable of acquiring highly effective non-uniform representations, it is important to note that in order to predict adjacent representations or reconstruct the original audio, the representation should encompass speaker-related information.

\setlength{\intextsep}{1.5mm}%
\setlength{\columnsep}{6pt}%
\begin{wrapfigure}{r}{0.5\linewidth}
    \vspace{-3mm}
    \centering
    \includegraphics[width=\linewidth]{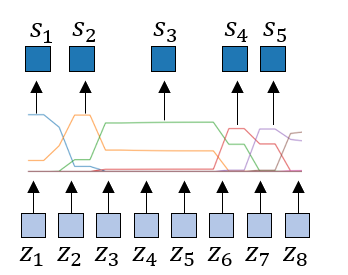}
    \vspace{-7mm}
    \caption{The concept of soft pooling}
    \label{fig: concept}
    \vspace{-2.7mm}
\end{wrapfigure}

Our objective is to minimize the presence of speaker-related information by augmenting the data through temporal misalignment and subsequently tasking the model with predicting the altered boundaries. 
This approach falls under the domain of contrastive learning through augmentation. What sets it apart is that the positive pair is not predefined; instead, it varies based on the model's predictions of the boundary. The overarching objective of this research is to enhance phonetic information while concurrently diminishing the presence of speaker-related details within speech representations.

\section{Related Works}
Creating positive samples by augmentation is a common method used in contrastive learning \cite{jiang2020speech, qian2022contentvec,lodagala2022ccc}.
The most augmentation methods for speech include pitch-shift, additive noise, applying room reverberation, or voice conversion. Time-stretch is hardly used because the position of a frame with the same information in augmented data is changed.
However, as time-stretch does not change the total number or order of phonemes, time-stretched samples can be used as a positive sample to embed linguistic information.
\cite{jiang2020speech} used speed perturbation not for linguistic information.
In singing domain, time-stretch was applied \cite{yakura2022self}, but augmented data was used as a negative sample.

This paper is related to ContentVec \cite{qian2022contentvec} in that it calculates the contrastive loss by creating positive samples by augmentation while maintaining the contents. \cite{qian2022contentvec} proposed three methods for disentanglement of speaker information in HuBERT framework: for student, teacher, and predictor. \cite{lodagala2022ccc} is an application of a similar idea in the wav2vec2.0 framework.
\cite{lodagala2022ccc} predicts a quantized latent representation as a context representation like wav2vec2.0, and yet the prediction target is a representation extracted from augmented data. The concept of contrastive learning through augmentation shares similarities with our approach, but our method deliberately introduces temporal misalignment along the time axis to predict boundaries.

Also, this paper has similarities to segmental CPC (SCPC) \cite{bhati2021segmental} and hierarchical CPC (hCPC) \cite{cuervo2022variable}, which predict the boundary and conduct pooling. In hCPC, the boundary is predicted to be 0 or 1 by sampling the action from the policy function. The policy function is learned to minimize the objective function of the high-level CPC. The differentiable pooling module in this paper is distinguishable in that it performs a weighted sum through attention rather than average pooling based on the boundary. Meanwhile, \cite{kreuk2020self} proposed a framework to detect phoneme boundaries in an unsupervised manner. Predicting a boundary with similarity between neighboring representations based on the CPC structure inspired this study. Our approach shares a fundamental idea with these methods, as we aim to predict the boundaries within the speech representation sequence. However, our method distinguishes itself by supplying positive pairs from augmented data rather than employing the same utterance. This unique approach allows us to determine boundaries based on content, thereby obviating the need for the inclusion of speaker-related information in the representation.

In this paper, we propose a self-supervised speech representation learning framework for optimizing the contrastive loss on augmented data, which is pooled based on the boundary.
Note that there is no supervision of the boundary and no text label. To the best of our knowledge, there was no self-supervised learning framework that adaptively pools time-stretched data  and compares it with the original data.

\begin{figure}[ht]
    \begin{center}
        \includegraphics[width=1.0\linewidth]{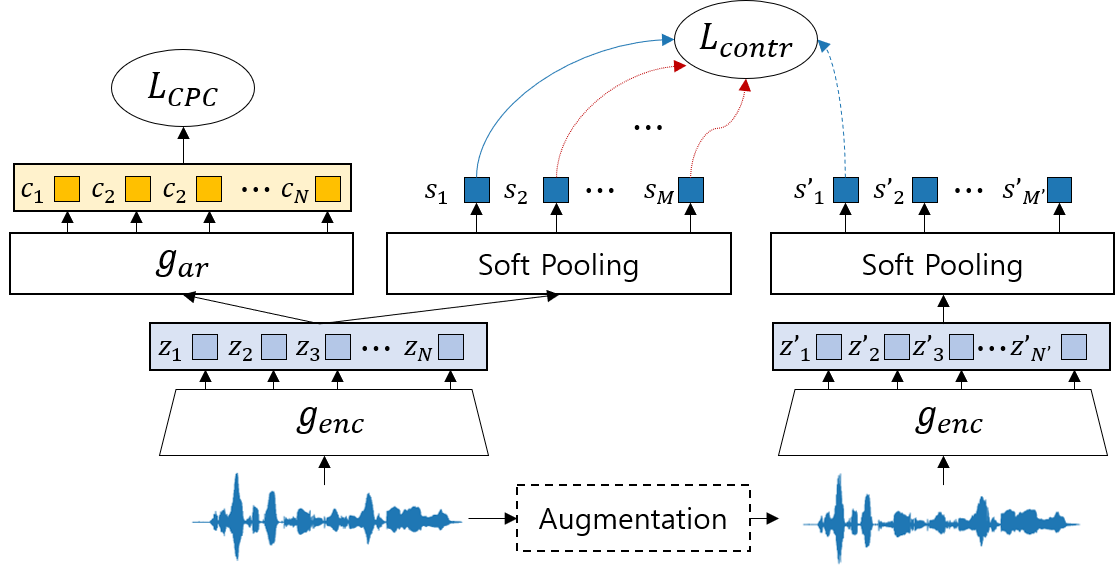}
    \end{center}
    \caption{Overall architecture of the model. $g_{enc}$, $g_{ar}$ are feature extractor and autoregressive network for extracting context vectors, respectively. Contrastive loss is calculated from the pooled representations of the original data and the augmented data. The blue solid line, the blue dotted line, and the red dotted line represent the anchor, positive sample, and negative sample, respectively.}
    \label{fig:model}
\end{figure}

%% file: 02_proposed_framework.tex
\section{Method}
\begin{figure}[ht]
    \begin{center}
        \includegraphics[width=0.4\linewidth]{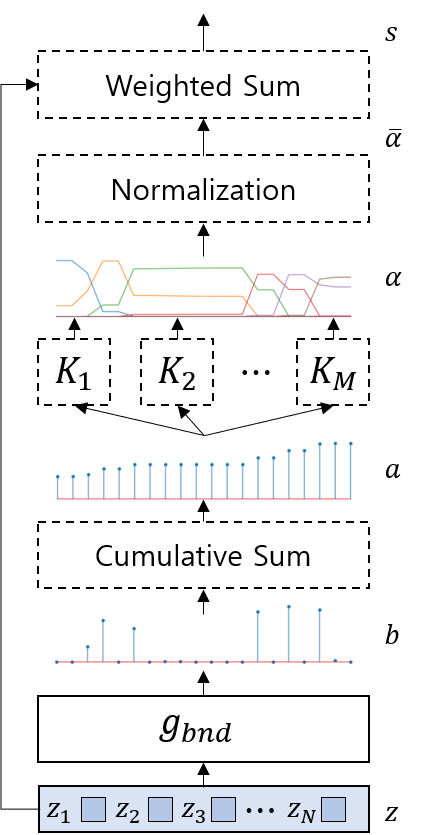}
    \end{center}
    \caption{Description of soft pooling module.}
    \label{fig:model}
\end{figure}

The overall architecture of proposed model and the soft pooling module are depicted in Figure \ref{fig:model}. The model is divided into two streams: a Contrastive Predictive Coding(CPC) part and a contrastive loss with augmentation part. Contrastive loss induces the boundary predictor to be based on linguistic information so that speech pairs with the same linguistic information are aligned on the time axis. The CPC loss is responsible for maintaining the linguistic information that can be lost while the contrastive loss is optimized.

\subsection{Contrastive Predictive Coding}

The procedure for obtaining the d-dimensional spectral representation $z$ from speech using convolution layers and the context vector $c$ through Long Short-Term Memory(LSTM) is identical to that employed in the CPC \cite{oord2018representation} framework. The encoders are trained by minimizing the contrastive loss$L_{CPC}$ calculated from $z$ and $c$
\begin{align}
L_{CPC} = -\sum_n \sum_{k=1}^{K}log\frac{f_k(c_n, z_{n+k})}{\Sigma_{z_i \in 
\mathcal{N}}f_k(c_n,z_i)},
\end{align}
where $f_k$ is prediction head for k upcoming representation. 
we opted to utilize CPC as a baseline and complement it with the incorporation of a soft pooling module.
\subsection{Augmentation}

In our augmentation process, we incorporated pitch-shift and time-stretch techniques that preserve the speech's underlying content.
When applying time-stretch, it should be noted that the positive samples are not time-aligned in the fixed-rate latent representation.
If variable-rate pooling reduces it to an event-based representation, the representation sequence will be aligned regardless of the length at which a particular linguistic unit is pronounced.
In essence, time-stretching prompts the boundary predictor to discern linguistic information boundaries.
To create a more challenging scenario for boundary prediction, we divided each speech sample into three segments, subjected them to random time-stretching, and subsequently concatenated them. Additionally, random pitch-shift was introduced to prevent boundary prediction based solely on pitch alterations.

\subsection{Soft Pooling Module}

We used attention pooling \cite{nadaraya1964estimating} to pool the representation sequence with predicted boundary probability. The probability is bounded from 0 to 1 using the sigmoid function as the activation function.
\begin{align}
    s_m = \sum_{n=1}^N\alpha_{mn}z_n=\sum_{n=1}^N\frac{K_m(a_n)}{\Sigma_{i=1}^N K_m(a_i)}z_n\\
    K_m(a_n) = \exp{(-0.5\sigma^{-2}(a_n - m)^2}),
\end{align}
where $N$ is the number of sequence length, $K_m$ is a Gaussian kernel for $m$-th attention head. As $a_n$ is cumulative sum of boundary, attention becomes monotonic with $m$.
The kernel causes $a_n$ to have more attention as $a_n$ is closer to $m$.
Ideally, if the boundaries only have values of 0 or 1 and the $\sigma$ is sufficiently reduced, attention would be equivalent to averaging pooling representations between non-zero boundaries. In other words, the clearer the boundary, the closer $a_n$ to an integer, and then reducing the $\sigma$ helps to create attention that reduces side effects. Because the boundary is not ideally trained, sufficient tolerance is required for the kernel, and in this paper, the sigma is set to 0.5.

The number of attention heads M to be set smaller than N is also an important hyperparameter.
$M$ needs to be set large enough to prevent the attention of $s_M$ from ending before reaching the end of the $z$ sequence.
Since the average number of phonemes per sec is about 10 and the frame rate of $g_{enc}$ is 100, $M = N/10$ can also be considered, but $M = N/4$ was conservatively set.
If the data is composed of few linguistic units (in the extreme case, silent audio), $s_m$ will be close to 0-vector for large m, which will not have much impact.

It is possible to substitute $c$ for $z$ used to calculate attention weight(key, query) or $z$ to be pooled(value). However, when $c$ is used as a key and query, the performance is slightly degraded, and when $c$ is used as a value, it collapses to predict the boundary as 1 for all frames. Thus, only $z$ is used in soft pooling module.

\subsection{Contrastive Loss}

Soft pooling module is trained by following objective function;
\begin{align}
L_{contr} = -\sum_{m=1}^{M}\log\frac{f(s_m,s'_m)}{\Sigma_{j=1}^{M}f(s_m,s_j)}\\
f(s^{(1)}, s^{(2)}) = \exp(\mathrm{sim}(s^{(1)},s^{(2)})/\tau),
\end{align}
where $\tau$ is temperature and sim is similarity function such as cosine similarity.
Positive sample is representations of the same location in the augmented data, and negative samples are representations of different locations in the original data like disentanglement in student of ContentVec \cite{qian2022contentvec}.
This loss induces the downsampled representations $s$, $s’$ to be similar frame-wise than different frames within the same $s$, making them event-based representations.

%% file: 03_experiment.tex
\section{Experiment}
\subsection{Experiment Setup}
Proposed models and baseline CPC were trained on LibriSpeech 960hours dataset \cite{librispeech}. Every model was trained with 2 RTX 3080ti GPUs and batch size is set to 32. We used RAdam optimizer with learning rate of 0.001. In order to see feasibility with limited resources, the representation is set to 128 dimensions, and accordingly, the number of the model parameters is 600K. The training took about 60 hours for 600,000 steps.

As this paper introduces a framework that combines both the CPC loss and the proposed contrastive loss, it is imperative to conduct a comparison with the established CPC approach. For this purpose, we will consider CPC with a reduced embedding dimension of 128 dimensions as our baseline. Additionally, we will assess models trained solely with the proposed contrastive loss, without the incorporation of CPC.

\subsection{Evaluation Methods}

\noindent
\textbf{Phonetic ABX task.} As one of the black box and zero-shot metrics provided by Zero-Resource Speech 2021 challenge\cite{dunbar21_interspeech}, it evaluates whether the model distinguishes phonemes of speech well. The evaluation data is a sample drawn from libri-light and is given as a triphone. Given the speech samples of A, B, and X, it determines whether X belongs to A or B based on the representation sequence extracted by the model. A and B are speech of the same speaker, and it is divided into within/across depending on whether X is the same speaker. Through the phonetic ABX error rate, it is possible to evaluate the extent to which the representation contains phonetic information.

\noindent
\textbf{Speaker Identification.} It was used to evaluate speaker disentanglement in ContentVec as a task to distinguish speakers in the SUPERB task. We used s3prl toolkit for SID evaluation \cite{mockingjay}. The lower the accuracy of speaker classification, the less information about the speaker in the representation.

\noindent
\textbf{Phoneme Segmentation.} TIMIT is often used as a benchmark for phoneme segmentation because it provides labels for phonemes and their boundaries. The precision, recall, and F1 scores are calculated by comparing the predicted boundary with the ground truth. Through this, it is possible to check how well the predicted boundary conforms to the phoneme boundary by learning with contrastive learning without any labels. Since most of the speech representation learning methods do not explicitly predict the boundary, only the results of hCPC are reported for comparison.

\newcolumntype{C}{>{$}c<{$}} 
\begin{table}
    \caption{Phonetic ABX error rate and speaker identification accuracy(SID). Lower SID accuracy is desirable.}
    \vspace{-0.5\baselineskip}
    \label{tab:abx_sid}
    \centering{
    \begin{tabular}{ccc}
    \hline\hline
        Model & ABX $\downarrow$ & SID $\downarrow$\\
         \hline
         ContentVec \cite{qian2022contentvec} & 5.13 & 37.7\\
         CPC baseline & 8.19 & 26.3\\
         Proposed model & 7.61 & 17.7\\
         Proposed model w/o CPC & 13.8 & 5.18\\
         \hline \hline
    \end{tabular}
    }
    \vspace{-3mm}
\end{table}

\begin{table}[h]
    \caption{Phoneme segmentation performance on TIMIT dataset. The metrics are calculated with 20 ms tolerance}
    \label{tab:segmentation}
    \vspace{-0.5\baselineskip}
    \centering{
    \resizebox{\linewidth}{!}{
    \begin{tabular}{ccccc}
    \hline\hline
        Model & Precision $\uparrow$ & Recall $\uparrow$ & F1 $\uparrow$ & R-val $\uparrow$\\
         \hline
         Proposed model & 73.31 & 75.01 & 74.15 & 77.79\\
         Proposed model w/o CPC & 71.38 & 70.74 & 71.04 & 75.32\\
         
         \hline \hline
    \end{tabular}
    }
    }
    \vspace{-3mm}
\end{table}

\subsection{Experiment Results}

The table \ref{tab:abx_sid} showed lower phonetic error rates and higher SID accuracy for the proposed model compared to the baseline CPC model. This highlights the effectiveness of the soft pooling module, optimized via contrastive loss, in disentangling speaker-related information while enhancing linguistic content embedding. However, when relying solely on contrastive loss (without CPC loss), SID accuracy decreased, indicating the importance of minimizing speaker-related data for optimizing the proposed contrastive loss. However, it is worth noting that without CPC loss, some phonetic information was sacrificed. In cases where both losses were jointly optimized, the performance in phonetic tasks improved while concurrently achieving speaker information disentanglement. Although the phonetic performance falls short of that achieved by the HuBERT-based ContentVec model, a substantial reduction in speaker-related information was accomplished.

The table \ref{tab:segmentation} indicates that the boundaries predicted by the proposed model align with phoneme boundaries. Remarkably, the segmentation F1 score reaches 74.15\%, even in the absence of constraints explicitly dictating that the predicted boundaries must coincide with phoneme boundaries. Similar to hCPC, employing a prior that incorporates phoneme rates as a regularization mechanism for boundary prediction has the potential to yield further improvements.

\begin{figure}[t]
    \includegraphics[width=\linewidth]{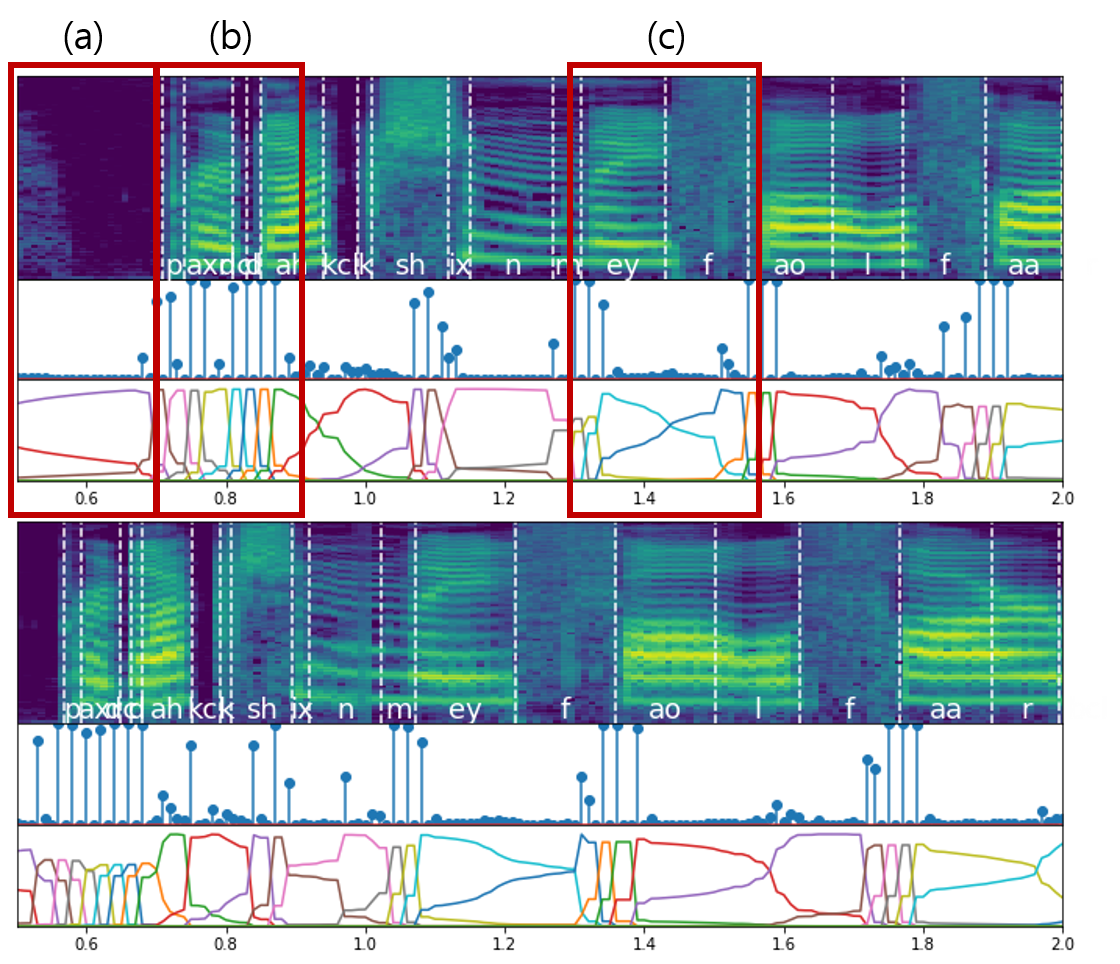}
    \caption{The three figures above are the mel-spectrogram with phoneme boundaries, predicted boundaries, and unnormalized attention weights of the original data, respectively, and the three figure below shows the same components for augmented data. Sample data is SI943 of FAKS0 from TIMIT.}
    \label{fig:example}
\end{figure}

The figure \ref{fig:example} displays the model's predicted boundaries and attention patterns. Although not all boundaries are precisely predicted, it's notable that regions with richer content engage more attention heads. For instance, the silent segment (a) and the extended /ey/ phoneme section (c) primarily involve one or two active attention heads, while the content-rich section (b) engages multiple attention heads. Despite variations in content distribution over time, both predicted boundaries and attention patterns share content-dependent trends.

